\documentclass[aps,prl,floatfix,twocolumn,showpacs,preprintnumbers,amsmath,amssymb]{revtex4-1}%

\usepackage{graphicx}
\usepackage{dcolumn}
\usepackage{bm}

\begin{document}


\title{Hydrodynamically-driven colloidal assembly in the thin-film entrainment regime}
\author{Carlos E. Colosqui}
\affiliation{Benjamin Levich Institute, 
City College of the City University of New York, New York, NY 10031, USA}
\author{Jeffrey F. Morris}
\affiliation{Benjamin Levich Institute, 
City College of the City University of New York, New York, NY 10031, USA}
\author{Howard A. Stone}
\email{hastone@princeton.edu} 
\affiliation{Department of Mechanical and Aerospace Engineering,
Princeton University, Princeton, NJ 08544, USA}
%
%
%
%
\begin{abstract}
We study numerically the hydrodynamics of dip coating from a suspension and report a mechanism for colloidal assembly and pattern formation on smooth and uniform substrates. 
Below a critical withdrawal speed of the substrate, capillary forces required to deform the meniscus prevent colloidal particles from entering the coating film.
Capillary forces are overcome by hydrodynamic drag only after a minimum number of particles organize in a close-packed formation within the meniscus. 
Once within the film, the formed assembly moves at nearly the withdrawal speed and rapidly separates from the next assembly.
The interplay between hydrodynamic and capillary forces can thus produce periodic and regular structures within the curved meniscus that extends below the withdrawn film.
The hydrodynamically-driven assembly documented here is consistent with stripe pattern formations observed experimentally in the so-called thin-film entrainment regime.
\end{abstract}
%
\maketitle
%
%
The ability to build crystalline microstructures on smooth  substrates enables significant advances in materials science and microfabrication, e.g. optoelectronics and metamaterials \cite{particles,Stebe2009}.
Film deposition from a colloidal suspension, via substrate withdrawal and/or solvent evaporation, can produce patterned microstructures that extend over millimeter lengths \cite{Giraldo2003,Abkarian2004,Stebe2007,Watanabe2009}.
For example, experimental studies have demonstrated a critical withdrawal speed below which the periodicity and regularity of the structures increase dramatically  \cite{Stebe2007,Watanabe2009}.
The critical speed corresponded closely to the conditions at which the particle diameter is equal to the thickness of a thin coating film predicted by the Landau-Levich theory \cite{Landau,*Wilson1982}.

These conditions set the upper limit of the so-called thin-film entrainment regime.
In this regime, the formation of periodic stripe patterns has been observed on partially wetting substrates \cite{Stebe2007} and further studied on hydrophilic substrates \cite{Watanabe2009}.
These experimental studies report that (i) very low particle concentrations are required for the formation of regular assemblies, and (ii) varying the withdrawal speed has a much greater effect on the stripe morphology than varying the solvent evaporation rate.
The employed technique, referred to as convective self-assembly (CSA), involves convective transport of colloidal particles from the bulk of the solution to the meniscus and coating film.
In regions where the assembly process takes place, the energies of convective motions and capillary interactions are much larger than the thermal energy $k_B T$ (here $k_B$ is Boltzmann's constant, and $T$ is temperature) and Brownian effects may be neglected.
Self-organization processes in CSA can thus involve nontrivial hydrodynamic interactions between colloidal particles.
Previous studies, however, attributed the stripe pattern formation to different mechanisms that only consider capillary interactions and quasi-static processes \cite{Stebe2007,Watanabe2009}.
In this Letter, we study and document the critical role of hydrodynamic effects in the particle assembly and subsequent pattern formation.

To study the hydrodynamics of the fluid-particle system, we employ a class of lattice Boltzmann (LB) methods for isothermal multiphase flow \cite{Shan1993,Colosqui2010}. 
The solid phase, which includes a moving wall (i.e. the withdrawn substrate), three stationary walls, and $N_p$ colloidal particles, is modeled with an immersed boundary (IB) approach \cite{Peskin2002,*Mittal2005}.
The employed IB approach represents the solid phase with a smooth distribution function, $\Phi_S({\bf x},t)\in [0,1]$ [see Fig.~1(a)] that determines local hydrodynamic forces due to (short-range) fluid-solid interactions (e.g. molecular collisions).
Long-range molecular interactions that give rise to capillary forces are modeled by two mean-field potentials, $\psi^{FF}({\bf x},t)$ for fluid-fluid and $\psi^{FS}({\bf x},t)$ for fluid-solid interactions \cite{Benzi2009,*Benzi2009b,Colosqui2012,*Kavousanakis2012}.
The interaction potential $\psi^{FF}$ models a volatile fluid that separates into two phases, which are modeled with an ideal equation of state \cite{Colosqui2012,*Kavousanakis2012}.
The coexistence densities are $\rho_L$~=~$m/(\Delta x)^3$ and $\rho_V$~=~0.1 $m/(\Delta x)^3$ for the liquid and vapor phase respectively; hereafter, $m$ is the molecular mass and $\Delta x$ is the numerical grid spacing.
The resulting surface tension is $\gamma$~$\simeq$~0.13~$k_B T/(\Delta x)^2$.
We adjust fluid-solid interactions so that the solid phase is always wetted by the liquid phase, which prevents colloidal particles from breaching the liquid-vapor interface.
For this work, we neglect long-range molecular interactions between solid bodies.
We also neglect thermal fluctuations that produce Brownian effects.
The modeled fluid-solid forces, ${\bf F}^{FS}\!\!$~=~${\cal F}\left[\Phi_S,\psi^{FF}\!\!,\psi^{FS}\right]$, govern the translational and rotational dynamics of the colloidal particles.
The equations of motion for the center-of-mass position, ${\bf x}_p(t)$, and angular velocity, $\boldsymbol{\omega}_p(t)$, of each particle $p$ are integrated using a conventional leapfrog scheme.
The numerical method is described in detail in the supplemental material.
%

The numerical model is applied to study the dynamics of colloidal assembly in the thin-film entrainment regime. 
Dip coating of plates produces a nearly two-dimensional flow that drives the convective transport of particles from the suspension bulk to the meniscus region and subsequently to the substrate.
Furthermore, the formation of regular stripes \cite{Giraldo2003,Stebe2007,Watanabe2009} requires a statistically homogeneous distribution of particles along the direction of the stripes.
We proceed by studying a two-dimensional flow configuration that resembles the classical Landau-Levich problem \cite{Landau,*Wilson1982} of plate coating via withdrawal from a liquid bath, where we later introduce circular particles. 
Notwithstanding these geometrical simplifications, we capture the basic mechanisms of particle assembly and stripe formation.

We perform numerical simulations in a rectangular domain that is partially bounded by walls.
The domain size is $L_x$~$\times$~$L_y$~$\simeq$~7~$\ell_c$~$\times$~3.2~$\ell_c$ where the capillary length is $\ell_c$~=~$\sqrt{\gamma/(\rho_L-\rho_V) g}$~$\simeq$~133~$\Delta x$ with a gravitational acceleration magnitude $g$~=~5 $\times 10^{-6}\!\Delta x/(\Delta t)^2$.
At initialization, the liquid bath lies below $y_H$~=~2.25~$\ell_c$ (see Fig.~1).
The vertical speed on the left wall, i.e. the withdrawal speed, is prescribed to small values, $U$~$\simeq$~0.25--2.0~$\times 10^{-2}\! \Delta x/ \Delta t$, which results in moderately low capillary numbers $Ca$~=~$U \mu / \gamma$~$ \simeq$~0.01--0.08.
Neutrally buoyant particles of circular shape and radius $R$~=~$10 \Delta x$ are used for the simulations in this Letter 
(the supplemental material includes results demonstrating grid convergence).
The particle dynamics are thus characterized by small Reynolds numbers $Re$~=~$\rho_L U R/ \mu$~$\simeq$~0.075--0.3 and a small Bond number $Bo$~=~$(R/\ell_c)^2$~=~6~$\times 10^{-3}$.
Particles in varying numbers, from zero to twenty, are released at the same time $t_0$ with zero initial velocity from different locations within the left half of the liquid bath; i.e. $x_p(t_0)$~$\in$~$(x_W$+$R$,$L_x/2]$ and $y_p(t_0)$~$\in$~$[y_H-20 R, y_H-3R]$ ($x_W$ indicates the surface of the coated plate).
Details of the simulation procedure are provided in the supplemental material.

First, we study the background flow topology. 
Simulations without particles reach steady state after a time $t_{S}$~=~$(L_y-y_H)/U$.
The steady-state flow topology, reported in Fig.~1, is typically observed in dip coating \cite{Scheid2010}.
%
\begin{figure}
\center
\includegraphics[angle=0,width=1.0\linewidth]{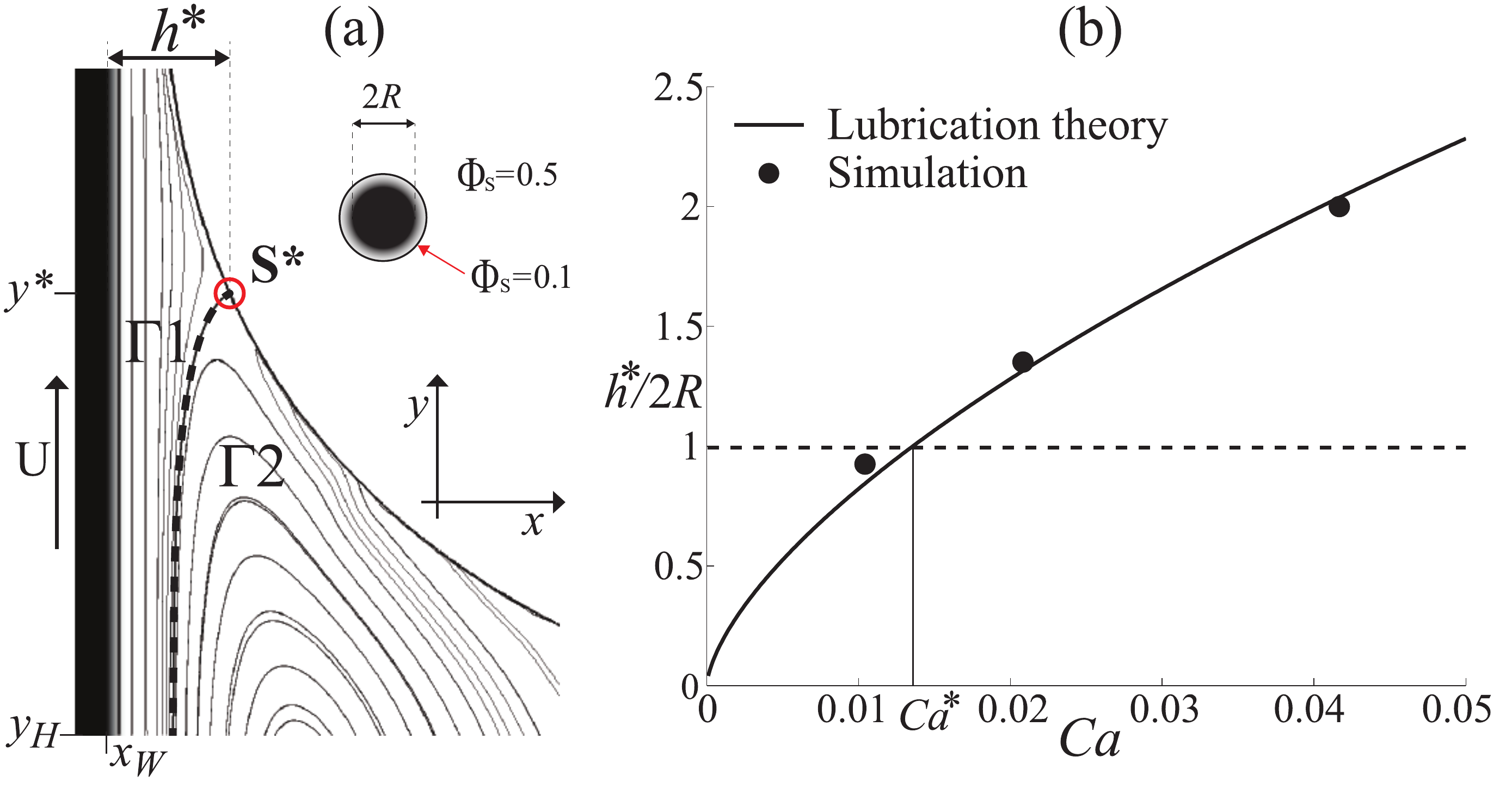}
\caption{Flow topology.
(a) Streamlines in the meniscus ($Ca$=0.04).
A stagnation point $\mathrm{S}^*$ (circle) lies on the interface where $h(y^*)$=$h^*$.
The streamline (dashed line) ending at $\mathrm{S}^*$ separates a ``shear flow'' region, $\Gamma 1$, from a ``recirculation flow'' region, $\Gamma 2$.
The surface of the withdrawn substrate is at $x$=$x_W$ and the horizontal level of the bath (at initialization) is at $y$=$y_H$.
A particle of size $R$=10$\Delta x$ is illustrated for reference; the diffuse fluid-solid interface is centered at $\Phi_S({\bf x},t)$=0.5.
(b) Normalized thickness $h^*/2 R$=$(h^{*}/\ell_c) / (2 \sqrt{Bo})$ vs. $Ca$=$U \mu/\gamma$.
Simulation results (symbols) compare well against lubrication theory (solid line) \cite{Landau,*Wilson1982}.
The critical entrainment condition $h^*$=$2 R$ corresponds to 
$Ca^*$$\simeq$$0.014$.
}
\label{fig1}
\end{figure}
%
A fundamental feature is the presence of a stagnation point $\mathrm{S}^*$ on the vapor-liquid interface at vertical position $y$~=~$y^*$ [cf. Fig.~1(a)].
The flow streamline ending at $\mathrm{S}^*$ defines the boundary between a ``shear flow'' region, $\Gamma 1$, which continues into the coating film, and a ``recirculation flow'' region, $\Gamma 2$, which extends into the liquid bath.
Under the studied conditions, the entrained film is always thinner than the particle diameter; one simulated particle is shown in Fig.~1(a) for reference.
The thickness profile $h(y)$ of the dynamic meniscus below the film largely determines whether a particle can enter the film.
Above the meniscus (for $y \gtrsim y_H+2 \ell_c$), the film thickness approximates the analytical prediction, 
$h_{f}/\ell_c$~=~0.95~$Ca^{2/3}-$~0.1~$Ca$, 
valid for $Ca\le 0.1$ \cite{Landau,*Wilson1982}.
More importantly, the meniscus thickness $h(y^*)=h^*$ observed at the stagnation point agrees well with predictions based on lubrication theory \cite{Landau,*Wilson1982}.
In Fig.~1(b) we report this characteristic thickness normalized by the particle diameter
$h^{*}/2R= [3 (h_{f}/\ell_c) - (h_{f}/\ell_c)^3/Ca] / 2 \sqrt{Bo}$.
There is a withdrawal speed corresponding to $Ca$~=~$Ca^*$~=~0.014 for which $h^{*}=2R$ [see Fig.~1(b)]; this condition appears to be critical for particle entrainment as we discuss below.

We now proceed to analyze the trajectories $\{x_p(t),y_p(t)\}$ after the background flow is established.  
For this purpose, we release particles simultaneously at $t_0$~=~1.2~$t_S$. 
We begin with the simplest case of a single particle, which yields a variety of open and closed trajectories.
Single particles released within the ``recirculation flow'' region $\Gamma 2$ attain stable orbits within the bath that prevent them from entering the film (see supplemental material).
The case of a single particle ($p$=I) released within the ``shear flow'' region $\Gamma 1$ is of particular interest. 
When approaching the film, the trajectories of single particles within $\Gamma 1$ are compressed into a unique ``entrainment'' trajectory by volume exclusion effects.
%
Consequently, when the withdrawal speed is sufficiently large, i.e. for $Ca>Ca^{*}$, a single particle successfully enters the film.
The vertical displacements $\Delta y$~=~$y_I(t)-y_H$~$\ge$~0 for different withdrawal speeds corresponding to $Ca$~=~0.02--0.08 are reported in Fig.~2(a).
%
\begin{figure}
\center
\includegraphics[angle=0,width=1.0\linewidth]{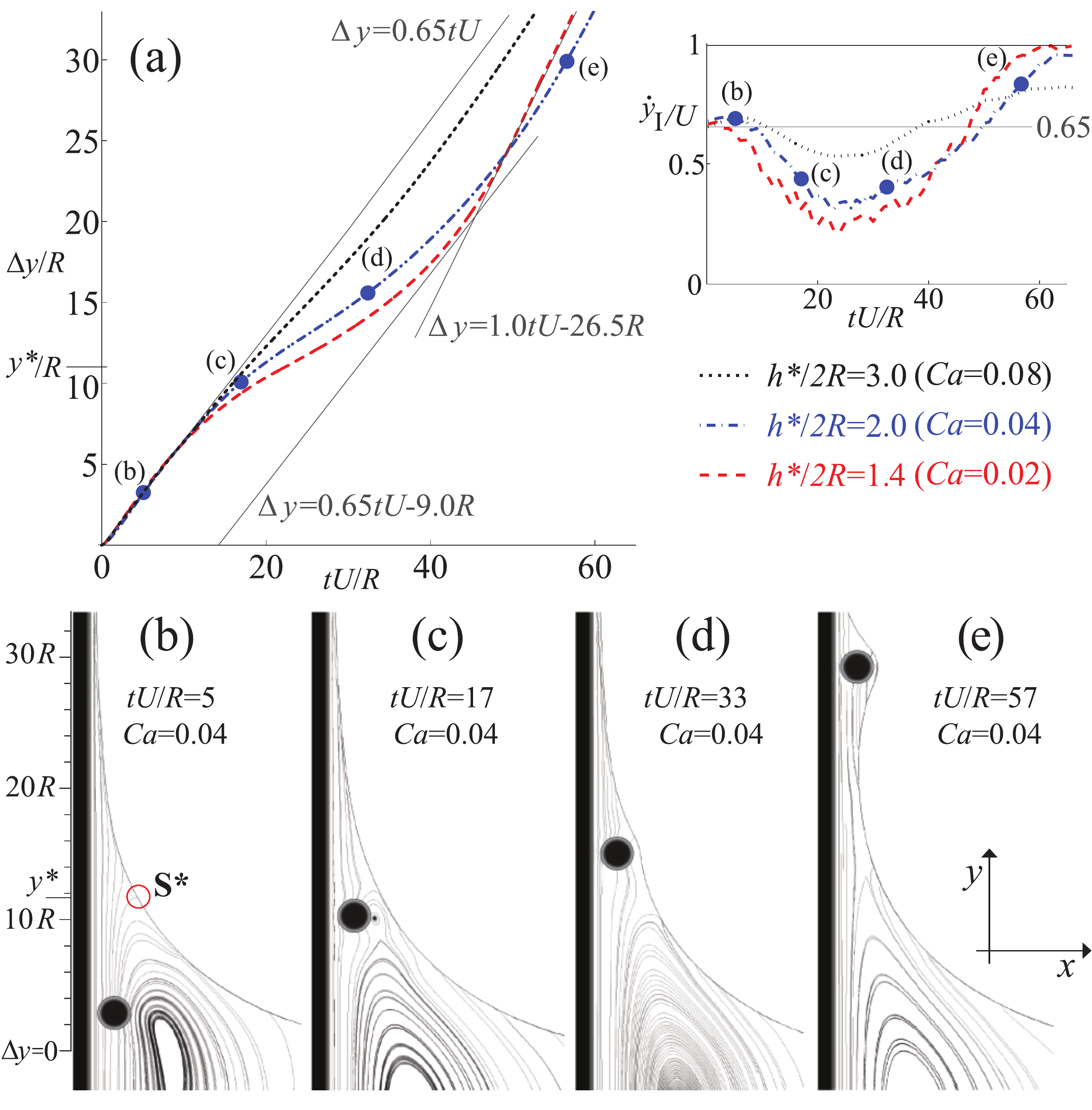}
\caption{Single particle dynamics for $Ca>Ca^*$.
(a) Vertical displacement $\Delta y=y_I(t)-y_H$ vs. $t U/R$ for $Ca>Ca^*$. 
The time origin ($t=0$) is chosen so that $\Delta y(0)=0$.
For $\Delta y>0$ individual particles released within $\Gamma 1$ follow a common $xy$-trajectory at a given $Ca$.
The inset in (a) shows the vertical speed $\dot{y}_I=dy_I/dt$.
(b--e) Streamlines for $Ca=0.04$ at different time instances indicated in panel (a).}
\label{fig2}
\end{figure}
%
A significant variation in the particle speed $\dot{y}=dy/dt$ can be observed in the inset of Fig.~2(a): 
the speed is nearly constant before entering the meniscus ($\dot{y}_I/U \simeq 0.65$ for $\Delta y/R<10$);
decreases inside the meniscus ($10\lesssim \Delta y/R \lesssim 20$);
and then increases after clearing the meniscus and entering the flat film ($\Delta y/R \gtrsim 20$).
The minimum vertical speed is reached near the stagnation point that is initially located at $y^*$.
While the minimum speed is directly correlated with the confinement ratio $h^*/2R$, the maximum vertical speed reached within the film is inversely correlated with $h^*/2R$.
The vertical speed is dominant along the entrainment trajectories in Fig.~2(a); i.e. $|\dot{x}_I|/|\dot{y}_I|\lesssim 0.1$ and $|\dot{w}_I R|/|\dot{y}_I|\lesssim 0.75$ ($\dot{w}_I$ is the angular velocity).

We next focus on the force balance along the nearly one-dimensional trajectory for entrainment.
As seen in Figs.~2(b--e), the entrainment of a single particle is accompanied by significant deformation of the meniscus and the flow streamlines.
In the horizontal direction, capillary forces are mainly balanced by a high pressure within the particle-wall gap and $\dot{x}_I \ll U$. 
In the vertical direction, hydrodynamic drag driving the particle motion is counteracted by capillary forces. 
For a confined particle moving parallel to the wall, the Stokes drag may be expressed as $F_{D}=-\mu (K_I \dot{y}_I-K_U U)$, where $K_I>0$ and $K_U>0$ are resistance coefficients.
The ratio of resistances $f_D=K_U/K_I\to 0$ for low confinement and far from the wall ($h/2R\gg 1$), while $f_D \sim 1$ in high confinement conditions and close to the wall ($h\sim 2R$). 
The capillary force (per unit length) is $F_{C}=-\gamma \partial A/\partial y_I$ where $A$ is the total interfacial area (per unit length).
For $Bo=(R/\ell_C)^2\ll 1$, we consider that capillary forces act normal to the interface with their magnitude determined by a shape function $f_C(h/2R)=|\partial A/ \partial y_I|$.
We then neglect inertial effects of order ${\cal O}(Re)$ to obtain
\begin{equation}
\label{eq:ydot}
\frac{\dot{y}_I}{U}= f_D(h/2R) + \frac{f_C(h/2R)}{K_I Ca} 
\frac{{\partial h}/{\partial y}}{\sqrt{1+|{\partial h}/{\partial y}|^2}}
\end{equation}
where $f_D$ and $f_C$ are positive functions determined by the thickness profile $h(y)$.
Hence, Eq.~(\ref{eq:ydot}) predicts a decay in vertical speed within the meniscus where $\partial h/\partial y<0$; this capillary effect increases for small $Ca$.
Once inside the film, we have $\partial h/ \partial y\simeq 0$ while the high confinement increases the drag, $f_D(h/2R)\to 1$, and thus the particle speed $\dot{y}_I\to U$.

%
Moreover, the force balance in Eq.~(\ref{eq:ydot}) predicts that colloidal particles can be trapped ($\dot{y_I}=0$) within the dynamic meniscus below a critical capillary number $Ca^*$.
%
%
Indeed, simulation results in Fig.~3 for a low withdrawal speed, where $Ca<Ca^*=0.014$ and $h^*<2R$, show that a single particle attains a stationary position within the meniscus.
%
\begin{figure}
\center
\includegraphics[angle=0,width=1.0\linewidth]{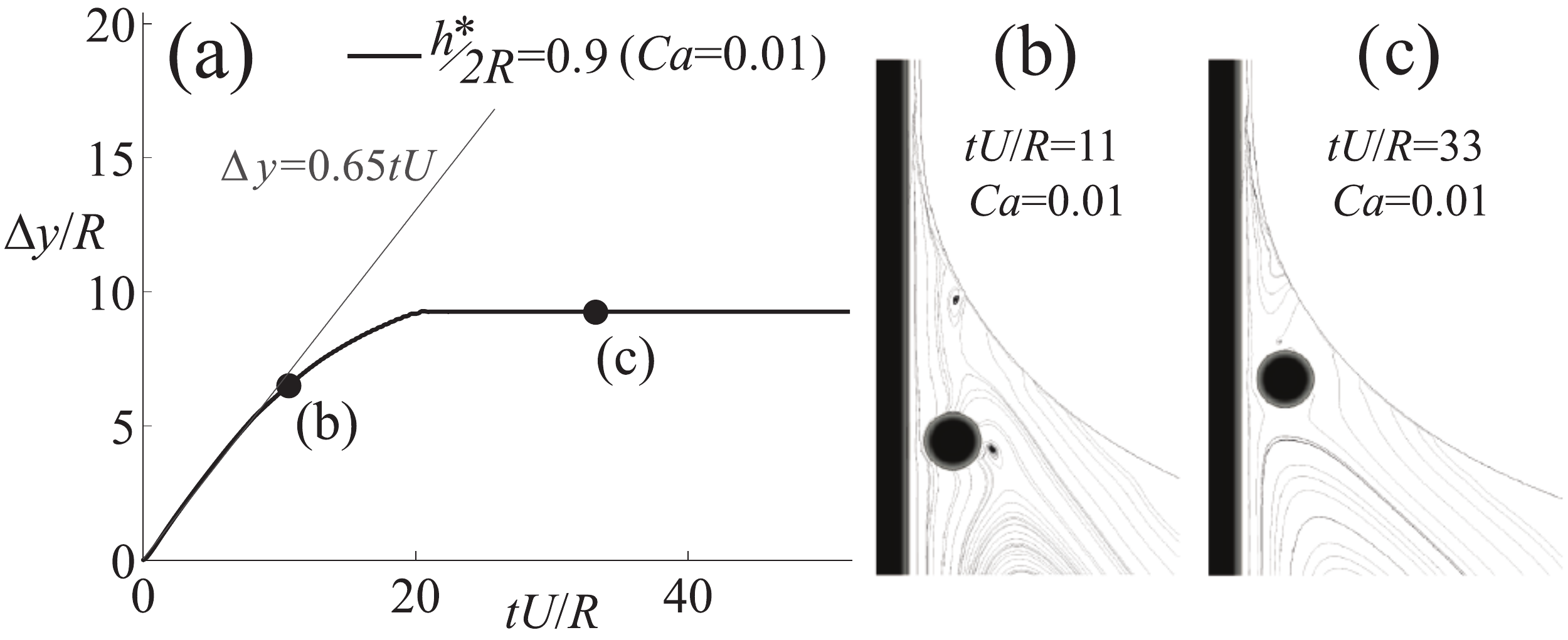}
\caption{Single particle dynamics for $Ca<Ca^*$.
(a) Vertical displacement $\Delta y=y_I(t)-y_H$ vs. $t U/R$. 
The time origin ($t=0$) is chosen so that $\Delta y(0)=0$.
Single particles released within $\Gamma 1$ reach a steady position inside the meniscus for $t U/ R$$>$25.
(b--c) Streamlines for $Ca=0.01<Ca^*$ at two time instances indicated in panel (a).
}
\label{fig3}
\end{figure}
%
Thus, Eq.~(\ref{eq:ydot}) indicates, through the functions $f_D$ and $f_C$, that widening and/or flattening the meniscus has critical effects on the particle speed.
For single particles we observe that $\dot{y}_I$~$\to$~0 for $h^*/2R$~$\to$~0, while $\dot{y}_I$~$\simeq$~const. for $h^*/2R$~$\to$~$\infty$ (cf. Figs.~2--3).
This observation suggests that a leading particle widens the meniscus, which would allow particles behind to climb faster.
This scenario is graphically illustrated by the slope and intercept of the straight lines in Fig.~2(a) for $h^*/2R=1.4$;
particles within $9R$ below a leading particle would assemble in the meniscus if moving with the speed ($\dot{y}$~$\simeq$~0.65~$U$) observed for $h^* \gtrsim 2R$. 
Moreover, particles already in the film would climb away about 26.5~$R$ during the assembly.
%

Simulations with multiple particles confirm a hydrodynamically-driven assembly within the meniscus for $Ca$~$=$~0.01~$<$~$Ca^*$.
%
\begin{figure}
\center
\includegraphics[angle=0,width=1.0\linewidth]{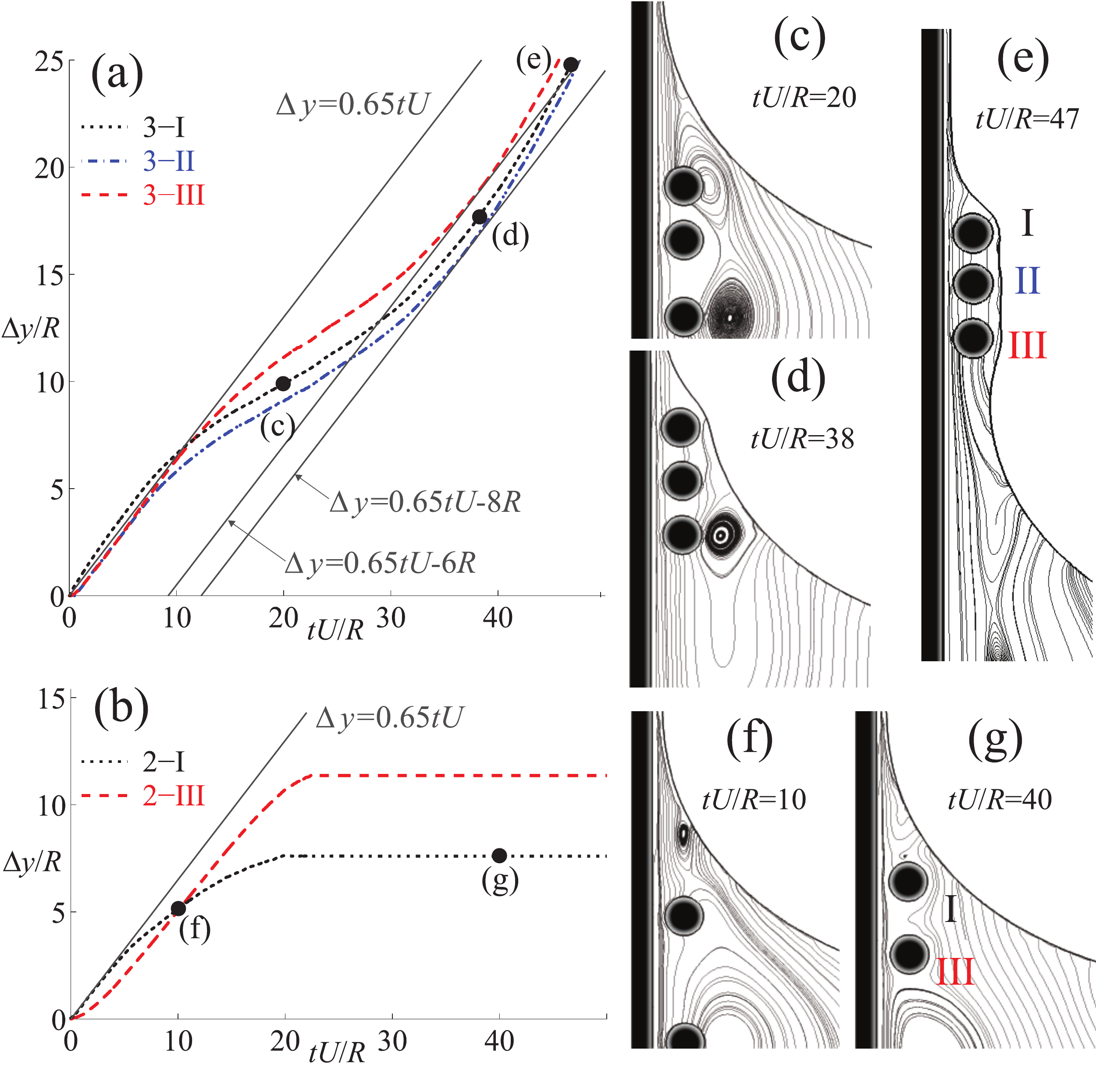}
\caption{Hydrodynamically-driven assembly for $Ca=0.01<Ca^*$.
(a--b) Vertical displacement $\Delta y_p=y_p(t)-y_p(0)$ vs. $t U/R$. 
Three ($p$=I-II-II) and two ($p$=I-II) particles are released at $t=0$; $y_{II}(0)=y_{I}(0)-3R$, $y_{III}(0)=y_{I}(0)-6R$, and $y_{I}(0)\simeq y_H$.
(c--e) and (f--g) Streamlines at time instances indicated in panels (a) and (b).
}
\label{fig4}
\end{figure}
%
Three particles ($p$=I-II-III) in one case [cf. Fig.~4(a)] and two particles ($p$=I-III) in a second case [cf. Fig.~4(b)] are released simultaneously with initial separations $y_{I}-y_{II}$~=~3$R$ and $y_{I}-y_{III}$~$=$~9$R$ where $y_I(0)$~$\simeq$~$y_H$. 
The vertical displacements in Fig.~4(a) show that the trailing particle (III) climbs faster than the leading particles (I-II), as hypothesized above.
The intercept of the straight lines in Fig.~4(a) estimates maximum initial separations of 6--8$R$ for which the assembly is possible.
In both cases the particles reduce their initial separations and assemble inside the meniscus [cf. Figs.~4(c--e) and Figs.~4(f--g)].
Below the critical condition $Ca<Ca^*$, a minimum number of colloidal particles (three in the studied conditions) must assemble in a close-packed formation before film entrainment is possible.
A close packing of particles increases the hydrodynamic drag and flattens the local interface curvature so that $\partial h/ \partial y\to 0$ for particles inside the assembly.
When hydrodynamic drag overcomes interfacial forces, the particle assembly enters the film and increases its speed as seen in Figs.~4(c--e).

Simulations with larger numbers of particles ($N_p=$10--20) further corroborate the periodic formation of regular assemblies when $Ca$~$<$~$Ca^*$ [cf. Fig.~5(a)].
%
\begin{figure}
\center
\includegraphics[angle=0,width=1.0\linewidth]{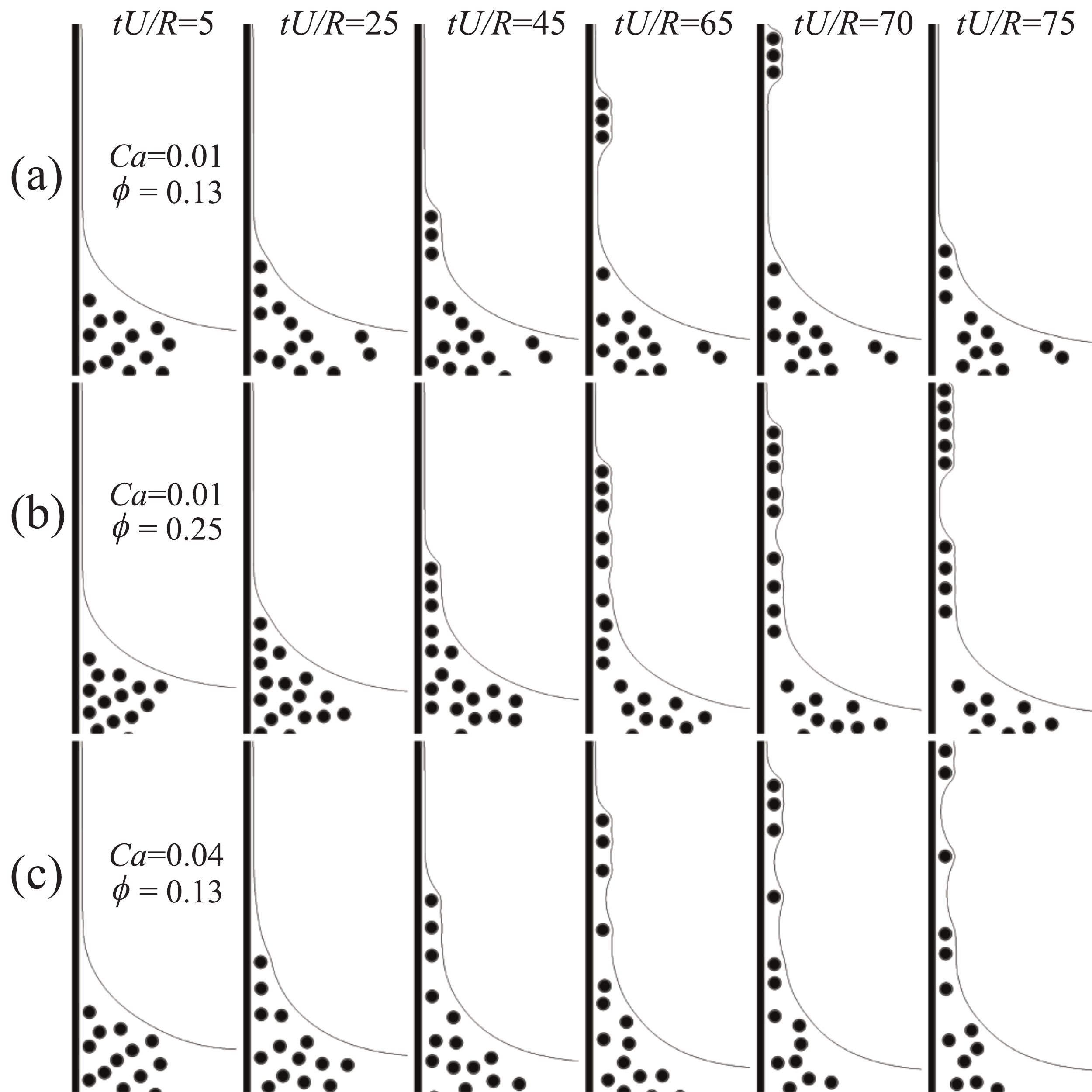}
\caption{Multiple particle assembly in thin films.
A sequence shows six time instances for 
(a) $Ca$=0.01 and $\phi$=$0.13R$, 
(b) $Ca$=0.01 and $\phi$=$0.25R$, and
(c) $Ca$=0.04 and $\phi$=$0.13R$.
For $Ca<Ca^*$ and low area factions $\phi\lesssim 0.13$, the assembly occurs in the dynamic meniscus, which increases the periodicity and regularity of the arrays (see movies in supplemental material).
}
\label{fig5}
\end{figure}
%
To draw this conclusion, we performed five realizations of each numerical experiment by releasing particles from random locations with root-mean-square particle separations $|\Delta {\bf x}_{rms}|=3.5 R$ and $|\Delta {\bf x}_{rms}|=2.5 R$, which corresponds to area fractions $\phi = 0.13$ and $\phi=0.25$ respectively.
The observed assembly within the dynamic meniscus was periodic and nearly identical (see movies included with the supplemental material).
For $Ca<Ca^*$ and the lower area fraction $\phi=0.13$ [cf. Fig.~5(a)], particle assemblies fully form before entering the film, which increases the periodicity and regularity of the arrays.
As seen in Figs.~5(b--c), increasing the area fraction or the withdrawal speed produces irregular assemblies that form once within the coating film. 
In the latter case, a different assembly mechanism, due to long-ranged capillary interactions and film instabilities \cite{particles}, deteriorates the periodicity and regularity of the observed stripe patterns.
%

%
In conclusion, our study indicates that below a critical withdrawal speed, for which $h^*$~$\lesssim$~$2R$, a hydrodynamically-driven assembly within the meniscus can produce highly organized structures.
The identified mechanism considers the coupling between hydrodynamic and capillary interactions.
Our simulations and analyses qualitatively agree with recent experimental observations of the formation of regular stripe patterns \cite{Stebe2007,Watanabe2009}.
The results in this Letter lead to a better fundamental understanding that could improve current technologies used for the flow-driven assembly of colloidal crystals.

%
CEC acknowledges stimulating discussions with M. Sbragaglia, X. Shan, and S. Succi.
CEC and JFM were supported by the NSF PREM (DMR-0934206).
HAS thanks the Princeton MRSEC for partial support.
%
%

\begin{thebibliography}{17}%
\makeatletter
\providecommand \@ifxundefined [1]{%
 \@ifx{#1\undefined}
}%
\providecommand \@ifnum [1]{%
 \ifnum #1\expandafter \@firstoftwo
 \else \expandafter \@secondoftwo
 \fi
}%
\providecommand \@ifx [1]{%
 \ifx #1\expandafter \@firstoftwo
 \else \expandafter \@secondoftwo
 \fi
}%
\providecommand \natexlab [1]{#1}%
\providecommand \enquote  [1]{``#1''}%
\providecommand \bibnamefont  [1]{#1}%
\providecommand \bibfnamefont [1]{#1}%
\providecommand \citenamefont [1]{#1}%
\providecommand \href@noop [0]{\@secondoftwo}%
\providecommand \href [0]{\begingroup \@sanitize@url \@href}%
\providecommand \@href[1]{\@@startlink{#1}\@@href}%
\providecommand \@@href[1]{\endgroup#1\@@endlink}%
\providecommand \@sanitize@url [0]{\catcode `\\12\catcode `\$12\catcode
  `\&12\catcode `\#12\catcode `\^12\catcode `\_12\catcode `\%12\relax}%
\providecommand \@@startlink[1]{}%
\providecommand \@@endlink[0]{}%
\providecommand \url  [0]{\begingroup\@sanitize@url \@url }%
\providecommand \@url [1]{\endgroup\@href {#1}{\urlprefix }}%
\providecommand \urlprefix  [0]{URL }%
\providecommand \Eprint [0]{\href }%
\providecommand \doibase [0]{http://dx.doi.org/}%
\providecommand \selectlanguage [0]{\@gobble}%
\providecommand \bibinfo  [0]{\@secondoftwo}%
\providecommand \bibfield  [0]{\@secondoftwo}%
\providecommand \translation [1]{[#1]}%
\providecommand \BibitemOpen [0]{}%
\providecommand \bibitemStop [0]{}%
\providecommand \bibitemNoStop [0]{.\EOS\space}%
\providecommand \EOS [0]{\spacefactor3000\relax}%
\providecommand \BibitemShut  [1]{\csname bibitem#1\endcsname}%
\let\auto@bib@innerbib\@empty
\bibitem [{\citenamefont {Kralchevsky}\ and\ \citenamefont
  {Nagayama}(2001)}]{particles}%
  \BibitemOpen
  \bibfield  {author} {\bibinfo {author} {\bibfnamefont {P.}~\bibnamefont
  {Kralchevsky}}\ and\ \bibinfo {author} {\bibfnamefont {K.}~\bibnamefont
  {Nagayama}},\ }\href@noop {} {\emph {\bibinfo {title} {Particles at Fluids
  Interfaces and Membranes}}},\ Vol.~\bibinfo {volume} {10}\ (\bibinfo
  {publisher} {Elsevier Science},\ \bibinfo {year} {2001})\BibitemShut
  {NoStop}%
\bibitem [{\citenamefont {Stebe}\ \emph {et~al.}(2009)\citenamefont {Stebe},
  \citenamefont {Lewandowski},\ and\ \citenamefont {Ghosh}}]{Stebe2009}%
  \BibitemOpen
  \bibfield  {author} {\bibinfo {author} {\bibfnamefont {K.}~\bibnamefont
  {Stebe}}, \bibinfo {author} {\bibfnamefont {E.}~\bibnamefont {Lewandowski}},
  \ and\ \bibinfo {author} {\bibfnamefont {M.}~\bibnamefont {Ghosh}},\
  }\href@noop {} {\bibfield  {journal} {\bibinfo  {journal} {Science}\ }\textbf
  {\bibinfo {volume} {325}},\ \bibinfo {pages} {159} (\bibinfo {year}
  {2009})}\BibitemShut {NoStop}%
\bibitem [{\citenamefont {Giraldo}\ \emph {et~al.}(2003)\citenamefont
  {Giraldo}, \citenamefont {Durand}, \citenamefont {Ramanan}, \citenamefont
  {Laubernds}, \citenamefont {Suib}, \citenamefont {Tsapatsis}, \citenamefont
  {Brock},\ and\ \citenamefont {Marquez}}]{Giraldo2003}%
  \BibitemOpen
  \bibfield  {author} {\bibinfo {author} {\bibfnamefont {O.}~\bibnamefont
  {Giraldo}}, \bibinfo {author} {\bibfnamefont {J.}~\bibnamefont {Durand}},
  \bibinfo {author} {\bibfnamefont {H.}~\bibnamefont {Ramanan}}, \bibinfo
  {author} {\bibfnamefont {K.}~\bibnamefont {Laubernds}}, \bibinfo {author}
  {\bibfnamefont {S.}~\bibnamefont {Suib}}, \bibinfo {author} {\bibfnamefont
  {M.}~\bibnamefont {Tsapatsis}}, \bibinfo {author} {\bibfnamefont
  {S.}~\bibnamefont {Brock}}, \ and\ \bibinfo {author} {\bibfnamefont
  {M.}~\bibnamefont {Marquez}},\ }\href@noop {} {\bibfield  {journal} {\bibinfo
   {journal} {Angew. Chem.}\ }\textbf {\bibinfo {volume} {115}},\ \bibinfo
  {pages} {3011} (\bibinfo {year} {2003})}\BibitemShut {NoStop}%
\bibitem [{\citenamefont {Abkarian}\ \emph {et~al.}(2004)\citenamefont
  {Abkarian}, \citenamefont {Nunes},\ and\ \citenamefont
  {Stone}}]{Abkarian2004}%
  \BibitemOpen
  \bibfield  {author} {\bibinfo {author} {\bibfnamefont {M.}~\bibnamefont
  {Abkarian}}, \bibinfo {author} {\bibfnamefont {J.}~\bibnamefont {Nunes}}, \
  and\ \bibinfo {author} {\bibfnamefont {H.A.}~\bibnamefont {Stone}},\
  }\href@noop {} {\bibfield  {journal} {\bibinfo  {journal} {J. Am. Chem.
  Soc.}\ }\textbf {\bibinfo {volume} {126}},\ \bibinfo {pages} {5978} (\bibinfo
  {year} {2004})}\BibitemShut {NoStop}%
\bibitem [{\citenamefont {Ghosh}\ \emph {et~al.}(2007)\citenamefont {Ghosh},
  \citenamefont {Fan},\ and\ \citenamefont {Stebe}}]{Stebe2007}%
  \BibitemOpen
  \bibfield  {author} {\bibinfo {author} {\bibfnamefont {M.}~\bibnamefont
  {Ghosh}}, \bibinfo {author} {\bibfnamefont {F.}~\bibnamefont {Fan}}, \ and\
  \bibinfo {author} {\bibfnamefont {K.}~\bibnamefont {Stebe}},\ }\href@noop {}
  {\bibfield  {journal} {\bibinfo  {journal} {Lang.}\ }\textbf {\bibinfo
  {volume} {23}},\ \bibinfo {pages} {2180} (\bibinfo {year}
  {2007})}\BibitemShut {NoStop}%
\bibitem [{\citenamefont {Watanabe}\ \emph {et~al.}(2009)\citenamefont
  {Watanabe}, \citenamefont {Inukai}, \citenamefont {Mizuta},\ and\
  \citenamefont {Miyahara}}]{Watanabe2009}%
  \BibitemOpen
  \bibfield  {author} {\bibinfo {author} {\bibfnamefont {S.}~\bibnamefont
  {Watanabe}}, \bibinfo {author} {\bibfnamefont {K.}~\bibnamefont {Inukai}},
  \bibinfo {author} {\bibfnamefont {S.}~\bibnamefont {Mizuta}}, \ and\ \bibinfo
  {author} {\bibfnamefont {M.}~\bibnamefont {Miyahara}},\ }\href@noop {}
  {\bibfield  {journal} {\bibinfo  {journal} {Lang.}\ }\textbf {\bibinfo
  {volume} {25}},\ \bibinfo {pages} {7287} (\bibinfo {year}
  {2009})}\BibitemShut {NoStop}%
\bibitem [{\citenamefont {Landau}\ and\ \citenamefont {Levich}(1942)}]{Landau}%
  \BibitemOpen
  \bibfield  {author} {\bibinfo {author} {\bibfnamefont {L.}~\bibnamefont
  {Landau}}\ and\ \bibinfo {author} {\bibfnamefont {B.}~\bibnamefont
  {Levich}},\ }\href@noop {} {\bibfield  {journal} {\bibinfo  {journal} {Acta
  Physicochim. URSS}\ }\textbf {\bibinfo {volume} {17}} (\bibinfo {year}
  {1942})}\BibitemShut {NoStop}%
\bibitem [{\citenamefont {Wilson}(1982)}]{Wilson1982}%
  \BibitemOpen
  \bibfield  {author} {\bibinfo {author} {\bibfnamefont {S.}~\bibnamefont
  {Wilson}},\ }\href@noop {} {\bibfield  {journal} {\bibinfo  {journal} {J.
  Eng. Math.}\ }\textbf {\bibinfo {volume} {16}},\ \bibinfo {pages} {209}
  (\bibinfo {year} {1982})}\BibitemShut {NoStop}%
\bibitem [{\citenamefont {Shan}\ and\ \citenamefont {Chen}(1993)}]{Shan1993}%
  \BibitemOpen
  \bibfield  {author} {\bibinfo {author} {\bibfnamefont {X.}~\bibnamefont
  {Shan}}\ and\ \bibinfo {author} {\bibfnamefont {H.}~\bibnamefont {Chen}},\
  }\href@noop {} {\bibfield  {journal} {\bibinfo  {journal} {Phys. Rev. E}\
  }\textbf {\bibinfo {volume} {47}},\ \bibinfo {pages} {1815} (\bibinfo {year}
  {1993})}\BibitemShut {NoStop}%
\bibitem [{\citenamefont {Colosqui}(2010)}]{Colosqui2010}%
  \BibitemOpen
  \bibfield  {author} {\bibinfo {author} {\bibfnamefont {C.}~\bibnamefont
  {Colosqui}},\ }\href@noop {} {\bibfield  {journal} {\bibinfo  {journal}
  {Phys. Rev. E.}\ }\textbf {\bibinfo {volume} {81}},\ \bibinfo {pages}
  {026702} (\bibinfo {year} {2010})}\BibitemShut {NoStop}%
\bibitem [{\citenamefont {Peskin}(2002)}]{Peskin2002}%
  \BibitemOpen
  \bibfield  {author} {\bibinfo {author} {\bibfnamefont {C.}~\bibnamefont
  {Peskin}},\ }\href@noop {} {\bibfield  {journal} {\bibinfo  {journal} {Acta
  Numer.}\ }\textbf {\bibinfo {volume} {11}},\ \bibinfo {pages} {479} (\bibinfo
  {year} {2002})}\BibitemShut {NoStop}%
\bibitem [{\citenamefont {Mittal}\ and\ \citenamefont
  {Iaccarino}(2005)}]{Mittal2005}%
  \BibitemOpen
  \bibfield  {author} {\bibinfo {author} {\bibfnamefont {R.}~\bibnamefont
  {Mittal}}\ and\ \bibinfo {author} {\bibfnamefont {G.}~\bibnamefont
  {Iaccarino}},\ }\href@noop {} {\bibfield  {journal} {\bibinfo  {journal}
  {Annu. Rev. Fluid Mech.}\ }\textbf {\bibinfo {volume} {37}},\ \bibinfo
  {pages} {239} (\bibinfo {year} {2005})}\BibitemShut {NoStop}%
\bibitem [{\citenamefont {Benzi}\ \emph
  {et~al.}(2009{\natexlab{a}})\citenamefont {Benzi}, \citenamefont {Chibbaro},\
  and\ \citenamefont {Succi}}]{Benzi2009}%
  \BibitemOpen
  \bibfield  {author} {\bibinfo {author} {\bibfnamefont {R.}~\bibnamefont
  {Benzi}}, \bibinfo {author} {\bibfnamefont {S.}~\bibnamefont {Chibbaro}}, \
  and\ \bibinfo {author} {\bibfnamefont {S.}~\bibnamefont {Succi}},\
  }\href@noop {} {\bibfield  {journal} {\bibinfo  {journal} {Phys. Rev. Lett.}\
  }\textbf {\bibinfo {volume} {102}},\ \bibinfo {pages} {26002} (\bibinfo
  {year} {2009}{\natexlab{a}})}\BibitemShut {NoStop}%
\bibitem [{\citenamefont {Benzi}\ \emph
  {et~al.}(2009{\natexlab{b}})\citenamefont {Benzi}, \citenamefont
  {Sbragaglia}, \citenamefont {Succi}, \citenamefont {Bernaschi},\ and\
  \citenamefont {Chibbaro}}]{Benzi2009b}%
  \BibitemOpen
  \bibfield  {author} {\bibinfo {author} {\bibfnamefont {R.}~\bibnamefont
  {Benzi}}, \bibinfo {author} {\bibfnamefont {M.}~\bibnamefont {Sbragaglia}},
  \bibinfo {author} {\bibfnamefont {S.}~\bibnamefont {Succi}}, \bibinfo
  {author} {\bibfnamefont {M.}~\bibnamefont {Bernaschi}}, \ and\ \bibinfo
  {author} {\bibfnamefont {S.}~\bibnamefont {Chibbaro}},\ }\href@noop {}
  {\bibfield  {journal} {\bibinfo  {journal} {J. Chem. Phys.}\ }\textbf
  {\bibinfo {volume} {131}},\ \bibinfo {pages} {104903} (\bibinfo {year}
  {2009}{\natexlab{b}})}\BibitemShut {NoStop}%
\bibitem [{\citenamefont {Colosqui}\ \emph {et~al.}(2012)\citenamefont
  {Colosqui}, \citenamefont {Falcucci}, \citenamefont {Ubertini},\ and\
  \citenamefont {Succi}}]{Colosqui2012}%
  \BibitemOpen
  \bibfield  {author} {\bibinfo {author} {\bibfnamefont {C.}~\bibnamefont
  {Colosqui}}, \bibinfo {author} {\bibfnamefont {G.}~\bibnamefont {Falcucci}},
  \bibinfo {author} {\bibfnamefont {S.}~\bibnamefont {Ubertini}}, \ and\
  \bibinfo {author} {\bibfnamefont {S.}~\bibnamefont {Succi}},\ }\href@noop {}
  {\bibfield  {journal} {\bibinfo  {journal} {Soft Matter}\ }\textbf {\bibinfo
  {volume} {8}},\ \bibinfo {pages} {3798} (\bibinfo {year} {2012})}\BibitemShut
  {NoStop}%
\bibitem [{\citenamefont {Kavousanakis}\ \emph {et~al.}(2012)\citenamefont
  {Kavousanakis}, \citenamefont {Colosqui}, \citenamefont {Kevrekidis},\ and\
  \citenamefont {Papathanasiou}}]{Kavousanakis2012}%
  \BibitemOpen
  \bibfield  {author} {\bibinfo {author} {\bibfnamefont {M.}~\bibnamefont
  {Kavousanakis}}, \bibinfo {author} {\bibfnamefont {C.}~\bibnamefont
  {Colosqui}}, \bibinfo {author} {\bibfnamefont {I.}~\bibnamefont
  {Kevrekidis}}, \ and\ \bibinfo {author} {\bibfnamefont {A.}~\bibnamefont
  {Papathanasiou}},\ }\href@noop {} {\bibfield  {journal} {\bibinfo  {journal}
  {Soft Matter}\ } (\bibinfo {year} {2012})}\BibitemShut {NoStop}%
\bibitem [{\citenamefont {Scheid}\ \emph {et~al.}(2010)\citenamefont {Scheid},
  \citenamefont {Delacotte}, \citenamefont {Dollet}, \citenamefont {Rio},
  \citenamefont {Restagno}, \citenamefont {Van~Nierop}, \citenamefont {Cantat},
  \citenamefont {Langevin},\ and\ \citenamefont {Stone}}]{Scheid2010}%
  \BibitemOpen
  \bibfield  {author} {\bibinfo {author} {\bibfnamefont {B.}~\bibnamefont
  {Scheid}}, \bibinfo {author} {\bibfnamefont {J.}~\bibnamefont {Delacotte}},
  \bibinfo {author} {\bibfnamefont {B.}~\bibnamefont {Dollet}}, \bibinfo
  {author} {\bibfnamefont {E.}~\bibnamefont {Rio}}, \bibinfo {author}
  {\bibfnamefont {F.}~\bibnamefont {Restagno}}, \bibinfo {author}
  {\bibfnamefont {E.}~\bibnamefont {Van~Nierop}}, \bibinfo {author}
  {\bibfnamefont {I.}~\bibnamefont {Cantat}}, \bibinfo {author} {\bibfnamefont
  {D.}~\bibnamefont {Langevin}}, \ and\ \bibinfo {author} {\bibfnamefont
  {H.A.}~\bibnamefont {Stone}},\ }\href@noop {} {\bibfield  {journal} {\bibinfo
  {journal} {Europhys. Lett}\ }\textbf {\bibinfo {volume} {90}},\ \bibinfo
  {pages} {24002} (\bibinfo {year} {2010})}\BibitemShut {NoStop}%
\end{thebibliography}
%
%

%
\end{document}